\documentclass{iau} 
\usepackage{graphicx}

\title[Chemistry of field halo stars and streams] 
      {Chemical abundances of field halo stars\\- Implications for the building blocks of the Milky Way - }

\author[Ishigaki, M. N.]   
{Miho N. Ishigaki $^1$
}

\affiliation{$^1$Tohoku University, Astronomical Institute, \\ 6-3   \\ email: {\tt miho@astr.tohoku.ac.jp} }

\pubyear{2019}
\volume{351}  
\jname{Star Clusters: From the Milky Way to the Early Universe}
\editors{A. Bragaglia, M.B. Davies, A. Sills \& E. Vesperini, eds.}
\begin{document}

\maketitle

\begin{abstract}

  I would like to review recent efforts of detailed chemical abundance measurements
  for field Milky Way halo stars.  Thanks to the advent of wide-field spectroscopic
  surveys up to a several kpc from the Sun, large samples of field
  halo stars with detailed chemical measurements are continuously expanding. Combination
  of the chemical information and full six dimensional phase-space information
  is now recognized as a powerful tool to identify cosmological accretion events that
  have built a sizable 
  fraction of the present-day stellar halo. Future observational prospects with
  wide-field spectroscopic surveys and theoretical prospects with supernova
  nucleosynthetic yields are also discussed.

\end{abstract}

\firstsection 
\section{Introduction}

Chemical abundances in the photosphere of ancient stars provide fossil records to
link field stars with their original birth places and thus 
serve as an essential tool to re-construct the merging history of
our Milky Way Galaxy (\cite[Freeman \& Bland-Hawthorn 2002]{freeman02}).
The stellar halo, which predominantly consists of old stellar populations, 
is a particularly interesting target  
because orbital velocities of the stars are largely preserved over the Galactic history
 due to the long dynamical time. This makes the Milky Way stellar halo an
ideal laboratory to test theories of galaxy formation and evolution
in the context of the currently standard $\Lambda$CDM cosmology (\cite[Bullock \& Johnston 2005]{bullock05};
\cite[Robertson et al. 2005]{robertson05}; \cite[Font et al. 2006]{font06}; \cite[Cooper et al. 2010]{cooper10})
based on spatial position, kinematics and chemistry of individual stars.

Since the stellar halo is an extremely diffuse component
where $\sim 10^9$ M$_\odot$ of stars are distributed in a large volume
over $\sim 150-200$ kpc from the Galactic center and the local density is much less than 1\%
(\cite[Juric et al. 2008]{juric08}), studies of chemical abundances in field halo stars
take significant advantages from wide-field spectroscopic
surveys. Large samples of candidate halo stars with low metallicity
have been built by prism spectroscopic surveys (the {\it HK survey} or the {\it Hamburg/ESO survey}). Their follow-up
high-resolution spectroscopy have revealed characteristic chemical abundances among
stars with [Fe/H] lower than $-3$
(\cite[Beers \& Christlieb 2005]{beers05} and
reference therein; Section \ref{sec:chem}).

More recently, wide-field low-to-medium resolution spectroscopic surveys such as
the Sloan Extension for Galactic Understanding and Exploration (SEGUE; \cite[Yanny et al. 2009]{yanny09}),
Radial Velocity Experiment (RAVE; 
\cite[Steinmetz et al. 2006]{steinmetz06}), the
Large Sky Area Multi-Object Fiber Spectroscopic Telescope (LAMOST) survey (\cite[Deng et al. 2012]{deng12}), and the
APOGEE survey (\cite[Majewski et al. 2017]{majewski17}) have
significantly improved survey volumes for 
stellar samples with line-of-sight velocity and chemical abundance
measurements. Wide-field photometric
surveys employing narrow-band filters sensitive to the
calcium H and K lines, such as SkyMapper Southern
Sky Surveys (\cite[Keller et al. 2007]{keller07}) and Pristine survey (\cite[Starkenburg et al. 2017]{starkenburg17}) have also been
successful in discovering the
most chemically pristine stars. High-resolution spectroscopic
follow-up observations are
accumulating a number of interesting chemical signatures in field halo stars
(e.g., \cite[Li et al. 2016]{li16}) that could potentially update the current understanding of the
chemical structure of our Galaxy and on the nature of the earliest
generations of stars in the Universe (\cite[Frebel \& Norris 2015]{frebel15})

The large spectroscopic data sets are particularly powerful when they
are combined with accurate parallax and proper motion data 
provided by satellite missions such as {\it Hipparcos} in studying the chemodynamical
structure and evolution of our Galaxy (see \cite[Feltzing \& Chiba 2013]{feltzing13} for a review of this subject). 
Furthermore, the measured chemical abundances provide further insights into
the stellar birth environment making use of Galactic chemical evolution
models (e.g. \cite[Kobayashi et al. 2006]{kobayashi06}) and with individual
supernova yield models (e.g., \cite[Woosley \& Weaver 1995]{woosley95};
\cite[Heger \& Woosley 2010]{heger10}; \cite[Nomoto, Kobayashi, \& Tominaga 2013]{nomoto13}).

In this article, I would like to focus on
how the chemical information help distinguishing the origin of individual
field halo stars in the following three different categories: 
(1) local halo stars that characterize the gross chemodynamical
structure of the halo (Section \ref{sec:global}), (2) halo stars with
full space motions measured and identified as candidate members of kinematically
coherent streams (Section \ref{sec:kinematics}), and (3) metal-poor
stars that show peculiar chemical abundance patterns (Section \ref{sec:chem}). Finally,
future prospects with on-going and planed large spectroscopic surveys
of the Milky Way are discussed (Section \ref{sec:future}).

\section{The global chemodynamical structure of the stellar halo\label{sec:global}}  

Thanks to the recent wide-field photometric surveys, our understanding of the
stellar distribution in the 
halo has been dramatically changed over the last few decades
(\cite[Ivezi\'{c} et al. 2012]{ivezic12}). It was clearly demonstrated that
the stellar halo is far from a  
smooth and static distribution of a single stellar population but it exhibits
various spatially coherent substructures.
While the spatial coherence of stars 
could be washed out in about 10 Gyrs, as a result of dynamical evolution
of the Galaxy, kinematics and chemical abundances are preserved for a longer
time scale and thus provide information on the 
early Galactic history (\cite[Helmi \& White 1999]{helmiwhite99}).

Dating back to the pioneering work by \cite{eggen62},
the strength of combining kinematics and chemistry of field halo stars
has been widely recognized (e.g. \cite[Norris \& Ryan 1989]{norris89},
\cite[Chiba \& Beers 2000]{chiba00}). One of the global nature of the stellar halo
revealed by full space motions and [Fe/H] of nearby field halo stars
is the presence of at least
two structural components, the inner and
outer halo populations (\cite[Carollo et al. 2007]{carollo07}).

Detailed elemental abundances in nearby halo stars have provided crucial
information on the nucleosynthesis and chemical evolution of the
early Universe. At [Fe/H]$\lesssim -2$, $\alpha$-element-to-iron
ratios ([(Mg, Si, Ca)/Fe]) have found to be enhanced by $\sim 0.4$ dex
relative to the solar values, and trend and scatter seen in
other elements have been interpreted as the dependence of Type II supernova yields
on progenitor metallicity or on explosion physics
(e.g., \cite[McWilliam et al. 1995]{mcwilliam95}; \cite[Ryan, Norris, \& Beers 1996]{ryan96};
\cite[Cayrel et al. 2004]{cayrel04}).
At higher [Fe/H], the enhancement of the $\alpha$-elements was thought to continue
up to [Fe/H]$\sim -1$, whose behavior was found to be different from the majority of
stars in dwarf spheroidal satellite galaxies with measured elemental
abundances (e.g., \cite[Venn et al. 2004]{venn04}).
Exception to the [$\alpha$/Fe]-[Fe/H] trends were also 
reported (e.g., \cite[Carney et al. 1997]{carney97}) in particular for
halo stars whose orbit reaches the outer Galactic halo (\cite[Nissen \& Schuster 1997]{nissen97}).

Subsequent studies analyzed a larger samples of stars with known kinematics (\cite[Fulbright 2000]{fulbright00};
\cite[Fulbright 2002]{fulbright02}; \cite[Stephens \& Boesgaard 2002]{stephens02}; 
\cite[Zhang et al. 2009]{zhang09}; \cite[Ishigaki et al. 2010]{ishigaki10}).  Among others,
\cite[Nissen \& Schuster (2010)]{nissen10} 
have expanded their original sample of \cite[Nissen \& Schuster (1997)]{nissen97} to $\sim 100$
stars carefully selected to
have similar stellar parameters (effective temperature, surface gravity and [Fe/H]) and have
full space motions.
They have shown that the sample of nearby halo stars can be separated into the
two chemically different populations; namely, high-$\alpha$ and low-$\alpha$
stars. In addition to the alpha elements, the two
populations show systematic difference in various elemental abundances
including C, Ni, Zn (\cite[Nissen \& Schuster 2011]{nissen11}) and in the neutron capture elements (\cite[Fishlock et al. 2017]{fishlock17}).

An intriguing question is whether the inner and the outer halo populations identified by
kinematics and [Fe/H] reported by \cite[Carollo et al. (2007)]{carollo07}
exactly correspond to the high-$\alpha$ and low-$\alpha$ components,
respectively, reported by \cite[Nissen \& Schuster (2010)]{nissen10}. \cite[Ishigaki et al. (2012)]{ishigaki12}
carried out high-resolution spectroscopic analyses of stars
from \cite[Chiba \& Beers (2000)]{chiba00} to study the difference in detailed
elemental abundances among those selected to have characteristic kinematics
of the thick disk, inner and outer halo stars reported by \cite[Carollo et al. (2007)]{carollo07}.
They have shown that the [Mg/Fe] ratios in stars kinematically
compatible with the inner and outer halo stars in  \cite[Carollo et al. (2007)]{carollo07}
show a decreasing trend with [Fe/H] at [Fe/H]$>-1.5$ on average. \cite[Ishigaki et al. (2013)]{ishigaki13}
further analyzed Fe-peak and neutron-capture elemental abundances and have
found that [Eu/Fe] ratios are higher for the stars with low [Mg/Fe], mostly comprised of
the outer halo stars, at [Fe/H]$>-1.5$.

The high-resolution spectroscopic survey by the APOGEE project
significantly increased the sample
size, thus providing an updated view on the nature of the two chemically distinct
components seen in the local halo stars. Based on elemental abundances of
$\sim 3200$ giant stars from the APOGEE data in the SDSS Data Release 12, 
\cite[Hawkins et al. (2015)]{hawkins15} identified
the $\alpha$-rich and the $\alpha$-poor sequences of stars
in the [$\alpha$/Fe]-[Fe/H] plane at
$-1.2<$[Fe/H]$<-0.55$. By analyzing 
chemical abundance trends with [Fe/H] for fourteen elements,
including CNO, $\alpha$, and
Fe-peak elements, they also
showed that the two sequences of stars are distinguished
by the abundance ratios of O, Mg, S, Al, C$+$N.
Furthermore, the two groups of stars show
systematically different kinematics identified in the
Galactic longitude ($l$) versus
Galactic rest-frame radial velocity (GRV) space, suggesting different
origins for the two groups. With the
updated chemical abundance estimates for an 
expanded sample of $\sim 62,000$ stars from 
SDSS Data Release 13, \cite[Hayes et al. (2018)]{hayes18}
confirmed the presence of the high and low-Mg sequences
similar to the previous findings.  The studies of halo stars either
with precision differential analysis (e.g., \cite[Nissen \& Schuster 2010]{nissen10}) or
with a large statistical sample (e.g., \cite[Hayes et al. 2018]{hayes}), 
demonstrate that, with homogeneously
measured various elemental abundances, the chemistry alone could be used to
separate different stellar populations for moderately 
metal-poor halo stars ($-2\lesssim$[Fe/H]$\lesssim -0.5$). The APOGEE data,
in particular, probe a much larger volume than 
previous high-resolution spectroscopic studies. Indeed,
\cite[Fern\'{a}ndez-Alvar et al. (2017)]{fernandez-alvar17} have
analyzed chemical abundances of $\sim 400$ stars in the range
$5<r<30$ kpc and demonstrate that the stars at $r>15$ kpc show
different trends in elemental abundance ratios ([X/Fe]) at [M/H]$>-1.1$. It  
has become clear that the chemical dichotomy seen in the
solar neighborhood is part of a global structure, extending
to at least up to several kpc from the Sun.

In summary, the key improvement in the last 20 years has been the recognition that
the chemistry of field halo stars is not represented by 
a homogeneous $\alpha$/Fe-enhancement over a wide [Fe/H] range but exhibits variation
depending on local space motions and/or Galactocentric distances.
 Whether this chemical
diversity corresponds to the global accretion events that are now
suggested to make up a large fraction of the local halo (\cite[Helmi et al. 2018]{helmi18};
\cite[Belokurov et al. 2018]{belokurov18})  
remains to be investigated in the next generation surveys.
The origin and the fraction of the high-$\alpha$ halo component remain elusive. 
It has been found that the high-$\alpha$ halo stars are
chemically indistinguishable from
the thick disk stars (\cite[Hawkins et al. 2015]{hawkins15}). 
Further studies incorporating all the six dimensional phase space coordinates
together with more detailed neutron capture elemental abundances
are needed to put more constraints on the origin of high- and low-$\alpha$
stars and their connection to the thick disk population. 
Another remaining question is nucleosynthetic origin of the chemical
difference between the high- and low-$\alpha$ populations. As has been
pointed out by previous studies (e.g., \cite[Nissen \& Schuster 1997]{nissen97};
\cite[Fishlock et al. 2017]{fishlock17}), an additional contribution of
elements from Type Ia supernovae to the gas initially enriched by
core-collapse supernovae does not fully explain the observed chemical difference.

\section{Chemistry of kinematically interesting stars \label{sec:kinematics}}

Along with the global structure,
substructures in kinematic spaces have been identified in the solar
 neighborhood 
(\cite[Helmi et al. 1999]{helmi99}; \cite[Chiba \& Beers 2000]{chiba00}; \cite[Arifyanto \& Fuchs 2006]{arifyanto06}; \cite[Dettbarn et al. 2007]{dettbarn07}; \cite[Kepley et al. 2007]{kepley07};
\cite[Klement et al. 2008]{klement08}; \cite[Klement et al. 2009]{klement09}; \cite[Smith et al. 2009]{smith09}; \cite[Smith 2016]{smith16} and \cite[Liang et al. 2018]{liang18} for recent reviews of this subject). 
These kinematically interesting halo stars, that are
often referred to as "kinematic streams" are considered to have originated
from accretion of dwarf galaxies or globular clusters to the Milky Way halo.
Chemistry has provided the most stringent test  to distinguish the origin of these streams.

One of the best known kinematic streams is the H99 stream, which
was identified by kinematics mostly provided by the Hipparcos satellite
(\cite[Helmi et al. 1999]{helmi99},\cite[Chiba \& Beers 2000]{chiba00}). 
\cite{roederer10} analyzed high-resolution spectra of
13 candidate member stars selected to have kinematics consistent with the H99 stream. It was shown that
the candidate member stars have a wide range of [Fe/H], which rules out the possibility that
the H99 is originated from a dissolved star cluster. Instead,
the [X/Fe] ratios are nearly homogeneous and their scatter is found to be small.
While the Galactic dwarf satellite galaxies that have metallicity similar
to the H99 stream show evolution in [X/Fe] with [Fe/H], the
abundance ratios of the H99 stars are nearly constant with [Fe/H] except for
neutron capture elements. No signature of chemical enrichment
by Type Ia supernovae or AGB stars are found. The observed
abundance pattern do not stand out compared to the bulk of
field halo stars that have similar [Fe/H].

Another well studied kinematic stream is KFR08, which was originally
discovered by \cite[Klement et al. (2008)]{klement08} based on velocities
from the RAVE survey.
Follow up studies have confirmed the presence of this stream
(\cite[Klement et al. 2009]{klement09}; \cite[Bobylev et al. 2010]{bobylev10})
based on independent data sets. The nature of the stream, however, remains
elusive due to the uncertainties in distances and kinematics
as well as a small number of candidate member stars
(\cite[Klement et al. 2011]{klement11}). \cite[Liu et al. (2015)]{liu15}
analyzed high-resolution spectra for 16 candidate member stars of
the KFR08 stream and estimated detailed elemental abundances of 14 elements.
They have found that the 16 stars have a scatter in [Fe/H] as large as 0.29 dex
and therefore, it is unlikely they originated from the same star cluster.
By quantifying similarity in chemical abundances among these stars by the method
proposed by \cite[Mitschang et al. (2013)]{mitschang13},  three of the 16 stars are found to
show [X/Fe] ratios close each other. On the other hand, their vertical
velocities ($W$) exhibit a large dispersion, which does not support the hypothesis that
the three chemically similar stars were formed in the same star cluster. Instead, the observed elemental
abundances as well as the estimated ages are similar to those seen in the
thick disk stars. \cite[Liu et al. (2015)]{liu15} therefore conclude that the KFR08 stream
is a kinematic stream that formed as a result of dynamical interactions
among the Galactic disk stars.

Recently, \cite[Zhao et al. (2018)]{zhao18} analyzed high-resolution spectra obtained by
Subaru/HDS for six candidate member stars of another kinematic stream, LAMOST-L1,
discovered by the LAMOST survey. They have found that
the six stars show a large [Fe/H] dispersion and only small
dispersion for [X/Fe]. The large [Fe/H] dispersion
is not reproduced if the member stars were formed in a same star cluster.
On average, the member stars
of LAMOST-L1 show lower [$\alpha$/Fe], [Na/Fe] and [Ni/Fe] ratios
and higher [Ba/Y] ratio compared to the bulk of the field halo  
stars that share a similar [Fe/H] (\cite[Suda et al. 2008]{suda08}). The amount of the
offset from the trend of field stars is similar to that found for
low-$\alpha$ stars as reported by \cite[Nissen \& Schuster (2010)]{nissen10}.
The direction of the offsets in [X/Fe] from the field stars is similar
to those reported for dwarf spheroidal galaxies, which might suggest
that they were originally born in a dwarf galaxy that was accreted
to the Milky Way in the past.

The origin of nearby field halo stars on highly retro-grade orbits has been
debated for a while. It has been proposed that the stars are tidal debris of
a galaxy which once hosted the $\omega$-Centauri ($\omega$ Cen) globular cluster (e.g., \cite[Mizutani et al. 2003]{mizutani03}).
\cite[Majewski et al. (2012)]{majewski12} analyzed
high-resolution ($R\sim 55,000$) spectra of giant stars within $\sim 5$ kpc
from the Sun which have been found to belong to a kinematic substructure
with a highly retro-grade orbit. It was found that the majority of these
stars show enhanced [Ba/Fe] ratios compared to the field halo stars
similar to those observed in $\omega$ Cen stars. This finding
suggests that these retro-grade stars are likely tidal debris of $\omega$ Cen itself.
With a large compilation of $\sim$800 literature abundance data in the SAGA database
(\cite[Suda et al. 2008]{suda08}) cross matched with Gaia DR2,
\cite[Matsuno et al. (2019)]{matsuno19} show that the 
highly retro-grade stars show a different trend in the [$\alpha$-Fe]-[Fe/H]
space from that seen among high-energy orbit stars.
These retrograde stars show a [$\alpha$/Fe]-[Fe/H] down turn, which is often
called "knee" at [Fe/H] $\sim 0.5$ dex lower than that
corresponding to the high-energy orbit stars. 
\cite[Myeong et al. (2019)]{myeong19} demonstrate that the
sequence seen in the chemical abundance plane in \cite[Matsuno et al. (2019)]{matsuno19}
is likely tidal debris of a dwarf galaxy, named "Sequoia" galaxy,
which is suggested to be the second-largest galaxy that has
contributed to the halo stars in the solar neighborhood.

These studies have clearly demonstrated that, when the identification with accurate kinematics are available,
detailed chemical information provided by high-resolution
spectroscopy is powerful in discriminating the origin of individual
substructures. It is often the case, however, that the chemical
differences from the bulk population are as small as typical observational
uncertainties. Therefore, the interpretation
is often limited by uncertainties in chemical abundance estimates, a small
sample size and contamination of field halo stars. 
Homogeneously analyzed high-resolution spectra that will be made available
with the ongoing Gaia-ESO survey (\cite[Gilmore et al. 2012]{gilmore12}),
GALAH survey (\cite[De Silva et al. 2015]{desilva15}) as well as WEAVE (\cite[Dalton et al. 2014]{dalton14}) in the future will alleviate these difficulty.

\section{Using chemistry to directly identify accreted stars\label{sec:chem}}

Stars that exhibit chemistry similar to those found in
dwarf spheroidal galaxies currently orbiting the Milky Way have been known
for a while, although they are relatively rare. 
The most remarkable classical example is the three stars,
BD$+80^{\circ} 245$ ([Fe/H]$=-2.07$), G 4-36 ([Fe/H]$=-1.94$), and CS 22966-043
([Fe/H]$=-1.91$).  \cite[Ivans et al. (2003)]{ivans03} carried out
a detailed chemical abundance analysis finding  extremely low  $\alpha$ (Mg, Si, and Ca)
-to-iron and [(Sr, Ba)/Fe] ratios with large variations in Fe-peak elements for the
three stars.

Thanks to the large spectroscopic surveys, candidate stars with
extreme chemical patterns are more efficiently found.
\cite[Xing et al. (2019)]{xing19} have analyzed one of candidate of stars that have
very low [$\alpha$/Fe] ratios identified by the LAMOST survey.
A follow-up spectroscopy with Subaru/HDS 
has confirmed that this star shows the [Mg/Fe] ratio of $-0.4$ at [Fe/H]$=-1.2$.
Such a low [Mg/Fe] ratio is unusual for the halo star with comparable
metallicity, while it is similar to stars in classical dwarf
spheroidal galaxies such as Ursa Minor. On the other hand, the
star shows a remarkable enhancement in r-process elements, with the abundance
pattern comparable to the solar-system r-process pattern.

\cite[Sakari et al. (2019)]{sakari19} reported
a low-$\alpha$ and mildly r-process enhanced
star, RAVE J093730.5-062655, originally identified
in the RAVE survey. \cite[Sakari et al. (2019)]{sakari19}
made detailed comparison of the observed abundances
with yield models of Type Ia supernovae
to investigate whether the abundances are explained by contribution of Fe
from Type Ia supernovae. Although the existing yield models of Type Ia supernovae
do not exactly reproduce all the observed elemental abundances, this
is more likely formed out of gas enriched with Fe from Type Ia supernovae.
Its retro-grade orbit clearly suggests that this star has come from an
accreted dwarf galaxy.

Both of
the low-$\alpha$ stars of \cite[Sakari et al. (2019)]{sakari19} and \cite[Xing et al. (2019)]{xing19} exhibit
enhancement of r-process elements that are more frequently seen among much lower [Fe/H] stars.
In fact there is a growing evidence that r-process
enhanced stars are originated from dwarf galaxies (\cite[Roederer et al. 2018]{roederer18}).

For the case of stars with metallicity lower than [Fe/H]$\sim -3$, that are collectively called extremely metal-poor
(EMP) stars, the observed abundances are generally believed to be the result of only one or a few
supernovae of the very first stars in the Universe (e.g., \cite[Audouze \& Silk 1995]{audouze95}).

A sign of stochastic chemical enrichment has been seen among EMP stars
as a large scatter in observed elemental abundance ratios. The most remarkable
feature is the presence of carbon enhanced stars that do not show enhancement in
s-process elements (CEMP-no; \cite[Yong et al. 2013]{yong13},
    \cite[Placco et al. 2014]{placco14} but see \cite[Norris \& Yong 2019]{norris19} for the effect of
    3D/NLTE effects on the Fe and C abundance measurements for EMP stars). Since the fraction of binary
    stars among the CEMP-no is not
    particularly high compared to normal EMP stars, it is
    unlikely their atmospheric composition was 
    modified by a binary mass transfer and thus
are thought to reflect the abundance of gas from which these stars formed.
The origin of the CEMP-no stars has been debated for a while. The proposed
scenarios include rotating massive first stars (\cite[Maeder et al. 2015]{maeder15}),
    faint supernovae (\cite[Umeda \& Nomoto 2003]{umeda03}, \cite[Iwamoto et al. 2005]{iwamoto05}),
    inhomogeneous metal-mixing (\cite[Hartwig \& Yoshida 2019]{hartwig-yoshida19}) or
    the result of the properties of dusts that were responsible for cooling
    the gas from which these stars have formed (\cite[Chiaki et al. 2017]{chiaki17}).
    As the detailed elemental abundances become available for EMP stars, it becomes
    clear that some fraction of CEMP-no stars also show enhancement of intermediate-mass
    elements, including Na, Mg, Al, or Si (\cite[Bonifacio et al. 2018]{bonifacio18},
    \cite[Aoki et al. 2018]{aoki18}). The diversity in other elemental
    abundance seen in CEMP-no stars suggests that multiple mechanisms are required to
    fully explain the carbon enhancement (\cite[Yoon et al. 2016]{yoon16}).

      Recent large statistical sampling of EMP stars have
    identified stars that show significantly lower [$\alpha$/Fe] ratios than
    the majority of halo stars with similar metallicities
    (\cite[Cohen et al. 2013]{cohen13};
    \cite[Caffau et al. 2013]{caffau13}; \cite[Bonifacio et al. 2018]{bonifacio18}).
    Unlike the low-$\alpha$ stars with [Fe/H]$\sim -1$ which are,
    at least in part, more likely related to the Type Ia enrichment, the origin of
    the EMP stars with sub solar [$\alpha$/Fe] ratios remain largely unknown.  
    \cite[Kobayashi et al. (2014)]{kobayashi14} proposed that
    the stars have been enriched by supernovae of
    low-mass first stars. It has also been demonstrated by
    \cite[Hartwig et al. (2018)]{hartwig18} that some of the low-$\alpha$ stars in the sample of
     \cite[Bonifacio et al. (2018)]{bonifacio18} are more likely to have been enriched by more than one supernova
     of the first stars.

These studies imply that the chemically peculiar EMP stars
have formed in the environment dominated by stochastic chemical enrichment.
Such characteristic patterns are frequently reported in ultra-faint dwarf galaxies
currently orbiting around the Milky Way (e.g., \cite[Koch et al. 2008]{koch08};
\cite[Tolstoy et al. 2009]{tolstoy09}; \cite[Salvadori et al. 2015]{salvadori15})
and some of the classical dwarf galaxies (e.g.\cite[Venn et al. 2012]{venn12}).

\section{Key questions for the future \label{sec:future}}

The observations of chemistry of field halo stars have yielded various
intriguing questions to be
addressed in the next generation observing facilities.
One of such questions would be 
how to quantify the relative contribution of substructures to the smooth halo component. 
Quantification of halo populations that have different birth places (e.g., in-situ, kicked-out, accreted)
is the central issue to test the Galaxy formation model as has been
addressed by \cite[Unavane et al. (1996)]{unavane96}. Detailed chemical information is essential
to make further progress in this issue since phase-space coordinates can be largely
washed out as the result of the dynamical evolution of the Galaxy.
A drawback of the chemical analysis is that observations tend to be
incomplete compared to the photometric sample and thus frequently 
suffer from selection bias. In this case it would be 
difficult to obtain a quantitative conclusion about the fraction of
stars with given chemistry in the whole stellar halo population.

Cosmological simulations incorporating the chemical evolution in the
building blocks of the Galaxy provide a powerful tool to quantify and interpret
the emerging chemical observations (e.g., \cite[Font et al. 2006]{font06}; \cite[Zolotov et al. 2010]{zolotov10};
\cite[Tissera et al. 2013]{tissera13}).
Techniques to compare observations with these theoretical
predictions have been investigated by e.g., \cite[Schlaufman et al. (2012)]{schlaufman12} and
\cite[Lee et al. (2015)]{lee15}.
These studies provide a step forward to make full use of
spectroscopic data from large surveys on-going and planed in the near future
such as WEAVE (\cite[Dalton et al. 2014]{dalton14}), 4MOST (\cite[de Jong et al. 2014]{dejong14}),
{\it Milky Way Mapper} survey planned as part of SDSS-V
(\cite[Kollmeier et al. 2017]{kollmeier17}), and the PFS (\cite[Takada et al. 2014]{takada14}).

For the theoretical side, some of the chemical signatures seen in field halo stars 
are likely connected to specific nucleosynthesis mechanisms
in supernovae of the earliest generation of stars (e.g., \cite[Ezzeddine et al. 2019]{ezzeddine19}).
Further investigations of theoretical yield models are
crucial to better understand the stellar birth environment. In fact,
it has been pointed out that the elemental abundances of the Sun are not
fully explained by neither traditional nor modern
core-collapse and Type Ia supernova yield models (\cite[Simionescu et al. 2019]{simionescu19}).

With increasingly large sample of high-resolution spectroscopic samples,
it would be interesting to compare the elemental abundance
patterns to grids of supernova yield models to obtain their
statistical properties
(\cite[Tominaga et al. 2014]{tominaga14}, \cite[Placco et al. 2015]{placco15},
\cite[Ishigaki et al. 2018]{ishigaki18}). These studies have
been used to investigate the possible origin of extremely metal-poor
stars in terms of the physical properties of the very first
generation of stars. It is still difficult to reproduce observed
abundances of all the key elements by any given supernova yield models.
This is partly due to the still unknown physical mechanism of
stellar evolution and supernova nucleosynthesis.

\section{Summary}

Important observational results on the chemistry of field halo stars
described in this article can be summarized as follows:

\begin{itemize}
\item The chemistry of nearby field halo stars
  with [Fe/H]$\gtrsim -1.5$ consist of at least two populations that are
  distinguished in the trend in [X/Fe]-[Fe/H] plane for $\alpha$-elements
  as well as several other elements. This is likely connected to the global
  structural components such as the dual halo structure 
  and hints at the formation of the Milky Way with accretions of dwarf galaxies. 
\item Some of the kinematic streams show characteristic abundance patterns that have
  helped distinguishing their birth places (dwarf galaxies, star clusters or
  the Galactic disk).
  \item Chemically interesting field halo stars at [Fe/H]$\gtrsim -3$ show characteristic
  chemical signature of an accreted dwarf galaxy which is likely, at least in part, connected to
  additional Fe enrichment by Type Ia supernovae. Among the lower [Fe/H] stars, scatter
  in elemental abundance ratios are
  prominent, particularly for light-to-intermediate mass elements,
  C, Mg, or Si, which could be a signature of the stochastic chemical enrichment
  in the early Universe as well as the properties of the earliest generation of stars in the Universe.

\end{itemize}

These observations have lead to a transition of our understanding
of the nearby halo stars from a traditional picture of a predominantly $\alpha$-enhanced
stellar population down to [Fe/H]$\sim -1$, that is distinct from
currently surviving dwarf satellite galaxies, to a new picture of
highly complex stellar populations in both kinematics and chemistry.
At the same time, these findings provide intriguing questions to be
answered in future observational and theoretical efforts.

\end{document}